\documentclass{article}%
\usepackage{amsmath}%
\usepackage{amsfonts}%
\usepackage{amssymb}%
\usepackage{graphicx}

\begin{document}

\author{VLADISLAV V. KRAVCHENKO\\Depto. de Telecomunicaciones\\SEPI, ESIME Zacatenco\\Instituto Polit\'{e}cnico Nacional\\Av.IPN S/N, C.P.07738, D.F.\\MEXICO\\vkravche@maya.esimez.ipn.mx}
\title{{\LARGE On the relation between the Maxwell system and the Dirac equation}}
\maketitle

\section{Introduction}

The relation between the two most important in mathematical physics first
order systems of partial differential equations is among those topics which
attract attention because of their general, even philosophical significance
but at the same time do not offer much for the solution of particular problems
concerning physical models. The Maxwell equations can be represented in a
Dirac like form in different ways (e.g., \cite{Greiner}, \cite{Imaeda},
\cite{KSim}). Solutions of Maxwell's system can be related to solutions of the
Dirac equation through some nonlinear equations (e.g., \cite{Vaz}).
Nevertheless, in spite of these significant efforts there remain some
important conceptual questions. For example, what is the meaning of this close
relation between the Maxwell system and the Dirac equation and how this
relation is connected with the wave-particle dualism. In the present article
we propose a simple equality involving the Dirac operator and the Maxwell
operators (in the sense which is explained below). This equality establishes a
direct connection between solutions of the two systems and moreover, we show
that it is valid when a quite natural relation between the frequency of the
electromagnetic wave and the energy of the Dirac particle is fulfilled. Our
analysis is based on the quaternionic form of the Dirac equation obtained in
\cite{Krbag} and on the quaternionic form of the Maxwell equations proposed in
\cite{Krdep} (see also \cite{KSbook}). In both cases the quaternionic
reformulations are completely equivalent to the traditional form of the Dirac
and Maxwell systems.

\section{Preliminaries}

The algebra of complex quaternions is denoted by $\mathbb{H}(\mathbb{C})$.
Each complex quaternion $a$ is of the form $a=\sum_{k=0}^{3}a_{k}i_{k}$ where
$\{a_{k}\}\subset\mathbb{C}$, $i_{0}$ is the unit and $\{i_{k}|\quad
k=1,2,3\}$ are the quaternionic imaginary units:%

\[
i_{0}^{2}=i_{0}=-i_{k}^{2};\;i_{0}i_{k}=i_{k}i_{0}=i_{k},\quad k=1,2,3;
\]%

\[
i_{1}i_{2}=-i_{2}i_{1}=i_{3};\;i_{2}i_{3}=-i_{3}i_{2}=i_{1};\;
\]%
\[
i_{3}i_{1}=-i_{1}i_{3}=i_{2}.
\]
The\ complex imaginary\ unit\ \ $i$\ commutes with \ $i_{k}$, $k=\overline
{0,3}$.

We will use the\ vector\ representation\ of complex quaternions:
$a=\operatorname*{Sc}(a)+\operatorname*{Vec}(a)$ ,\ where\ $\operatorname*{Sc}%
(a)=a_{0}$ and $\operatorname*{Vec}(a)=\overrightarrow{a}\mathbf{=}\sum
_{k=1}^{3}a_{k}i_{k}$. That is each complex quaternion is a sum of its scalar
part and its vector part. Complex vectors we identify with complex quaternions
whose scalar part is equal to zero. In vector terms, the multiplication of two
arbitrary complex\ quaternions\ $a$ and $b$\ can\ be\ written\ as\ follows:\ %

\[
a\cdot b=a_{0}b_{0}-<\overrightarrow{a},\overrightarrow{b}>+\left[
\overrightarrow{a}\times\overrightarrow{b}\right]  +a_{0}\overrightarrow
{b}+b_{0}\overrightarrow{a},
\]

\noindent where\ %

\[
<\overrightarrow{a},\overrightarrow{b}>:={\displaystyle    \sum_{k=1}^{3}%
}a_{k}b_{k}\in\mathbb{C}
\]
and%

\[
\lbrack\overrightarrow{a}\times\overrightarrow{b}]:=\left|
\begin{array}
[c]{lll}%
i_{1} & i_{2} & i_{3}\\
a_{1} & a_{2} & a_{3}\\
b_{1} & b_{2} & b_{3}%
\end{array}
\right|  \in\mathbb{C}^{3}.
\]

We shall consider continuously differentiable $\mathbb{H(C)-}$valued functions
depending on three real variables $x=(x_{1},x_{2},x_{3})$. On this set the
well known (see, e.g., \cite{BDS}, \cite{GS1}, \cite{KSbook})
Moisil-Theodoresco operator is defined by the expression
\[
D:=\sum_{k=1}^{3}i_{k}\partial_{k},\qquad\text{where \qquad}\partial
_{k}=\frac{\partial}{\partial x_{k}}.
\]
The action of the operator $D$ on an $\mathbb{H(C)-}$valued function $f$ can
be written in a vector form:%

\begin{equation}
Df=-\operatorname*{div}\overrightarrow{f}+\operatorname*{grad}f_{0}%
+\operatorname*{rot}\overrightarrow{f}. \label{Dvec}%
\end{equation}
That is, $\operatorname*{Sc}(Df)=-\operatorname*{div}\overrightarrow{f}$ and
$\operatorname*{Vec}(Df)=\operatorname*{grad}f_{0}+\operatorname*{rot}%
\overrightarrow{f}$. In a good number of physical applications (see \cite{GS1}
and \cite{KSbook}) the operators $D_{\alpha}=D+M^{\alpha}$ and $D_{-\alpha
}=D-M^{\alpha}$ are needed, where $\alpha$ is a complex quaternion and
$M^{\alpha}$ denotes the operator of multiplication by $\alpha$ from the
right-hand side: $M^{\alpha}f=f\cdot\alpha$. Here we will be interested in two
special cases when $\alpha$ is a scalar, that is $\alpha=\alpha_{0}$ or when
$\alpha$ is a vector $\alpha=\overrightarrow{\alpha}$. The first case
corresponds to the Maxwell equations and the second to the Dirac equation.

\section{The Dirac equation}

Consider the Dirac equation in its covariant form%

\[
(\hbar(\frac{\gamma_{0}}{c}\partial_{t}+\sum_{k=1}^{3}\gamma_{k}\partial
_{k})+imc)\Phi(t,x)=0.
\]
For a wave function with a given energy we have $\Phi
(t,x)=q(x)e^{i\frac{\mathcal{E}}{\hbar}t}$, where $q$ satisfies the equation%
\begin{equation}
(\frac{i\mathcal{E}}{c\hbar}\gamma_{0}+\sum_{k=1}^{3}\gamma_{k}\partial
_{k}+\frac{imc}{\hbar})q(x)=0. \label{Diract}%
\end{equation}
Denote
\[
\mathbb{D}:=\frac{i\mathcal{E}}{c\hbar}\gamma_{0}+\sum_{k=1}^{3}\gamma
_{k}\partial_{k}+\frac{imc}{\hbar}.
\]

Let us introduce an auxiliary notation $\widetilde{f}:=f(t,x_{1},x_{2}%
,-x_{3})$. The transformation which allows us to rewrite the Dirac equation in
a quaternionic form we denote as $\mathcal{A}$ and define in the following way
\cite{Krbag}. A function $\Phi:\mathbb{R}^{3}\rightarrow\mathbb{C}^{4}$ is
transformed into a function $F:\mathbb{R}^{3}\rightarrow\mathbb{H(C)}$ by the
rule
\begin{align*}
F  &  =\mathcal{A}[\Phi]=\frac{1}{2}(-(\widetilde{\Phi}_{1}-\widetilde{\Phi
}_{2})i_{0}+i(\widetilde{\Phi}_{0}-\widetilde{\Phi}_{3})i_{1}-\\
&  (\widetilde{\Phi}_{0}+\widetilde{\Phi}_{3})i_{2}+i(\widetilde{\Phi}%
_{1}+\widetilde{\Phi}_{2})i_{3}).
\end{align*}
The inverse transformation $\mathcal{A}^{-1}$ is defined as follows
\[
\Phi=\mathcal{A}^{-1}[F]=
\]%
\[
(-i\widetilde{F}_{1}-\widetilde{F}_{2},-\widetilde{F}_{0}-i\widetilde{F}%
_{3},\widetilde{F}_{0}-i\widetilde{F}_{3},i\widetilde{F}_{1}-\widetilde{F}%
_{2}).
\]
Let us present the introduced transformations in a more explicit matrix form
which relates the components of a $\mathbb{C}^{4}$-valued function $\ \Phi$
with the components of an $\mathbb{H(C)}$-valued function $F$:
\[
F=\mathcal{A}[\Phi]=\frac{1}{2}\left(
\begin{array}
[c]{rrrr}%
0 & -1 & 1 & 0\\
i & 0 & 0 & -i\\
-1 & 0 & 0 & -1\\
0 & i & i & 0
\end{array}
\right)  \left(
\begin{array}
[c]{c}%
\widetilde{\Phi}_{0}\\
\widetilde{\Phi}_{1}\\
\widetilde{\Phi}_{2}\\
\widetilde{\Phi}_{3}%
\end{array}
\right)
\]
and%

\[
\Phi=\mathcal{A}^{-1}[{F}]=\left(
\begin{array}
[c]{rrrr}%
0 & -i & -1 & 0\\
-1 & 0 & 0 & -i\\
1 & 0 & 0 & -i\\
0 & i & -1 & 0
\end{array}
\right)  \left(
\begin{array}
[c]{c}%
\widetilde{F}_{0}\\
\widetilde{F}_{1}\\
\widetilde{F}_{2}\\
\widetilde{F}_{3}%
\end{array}
\right)  .
\]

We have the following important equality%
\begin{equation}
D_{\overrightarrow{\alpha}}=-\mathcal{A}\gamma_{1}\gamma_{2}\gamma
_{3}\mathbb{D}\mathcal{A}^{-1}, \label{svyaz2}%
\end{equation}
where $\overrightarrow{\alpha}:=-\frac{1}{\hbar}(i\frac{\mathcal{E}}{c}%
i_{1}+mci_{2})$. This equality shows us that instead of equation
(\ref{Diract}) we can consider the equation%
\begin{equation}
D_{\overrightarrow{\alpha}}f=0 \label{Dirac}%
\end{equation}
and the relation between solutions of (\ref{Diract}) and (\ref{Dirac}) is
established by means of the invertible transformation $\mathcal{A}$:
$f=\mathcal{A}q$.

\section{The Maxwell equations}

We will consider the time-harmonic Maxwell equations for a sourceless
isotropic homogeneous medium%
\begin{equation}
\operatorname*{rot}\overrightarrow{H}=-i\omega\varepsilon\overrightarrow{E},
\label{Mh1}%
\end{equation}%
\begin{equation}
\operatorname*{rot}\overrightarrow{E}=i\omega\mu\overrightarrow{H},
\label{Mh2}%
\end{equation}%
\begin{equation}
\operatorname*{div}\overrightarrow{E}=0, \label{Mh3}%
\end{equation}%
\begin{equation}
\operatorname*{div}\overrightarrow{H}=0. \label{Mh4}%
\end{equation}
Here $\omega$ is the frequency, $\varepsilon$ and $\mu$ are the absolute
permittivity and permeability respectively. $\varepsilon=\varepsilon
_{0}\varepsilon_{r}$ and $\mu=\mu_{0}\mu_{r}$, where $\varepsilon_{0}$ and
$\mu_{0}$ are the corresponding parameters of a vacuum and $\varepsilon_{r}$,
$\mu_{r}$ are the relative permittivity and permeability of a medium.

Taking into account (\ref{Dvec}) we can rewrite this system as follows
\begin{equation}
D\overrightarrow{E}=i\omega\mu\overrightarrow{H}, \label{DMax1}%
\end{equation}%
\begin{equation}
D\overrightarrow{H}=-i\omega\varepsilon\overrightarrow{E}. \label{DMax2}%
\end{equation}
This pair of equations can be diagonalized in the following way \cite{Krdep}
(see also \cite{KSbook}). Denote
\begin{equation}
\overrightarrow{\varphi}:=-i\omega\varepsilon\overrightarrow{E}+\kappa
\overrightarrow{H} \label{ph}%
\end{equation}
and
\begin{equation}
\overrightarrow{\psi}:=i\omega\varepsilon\overrightarrow{E}+\kappa
\overrightarrow{H}, \label{ps}%
\end{equation}
where $\kappa:=\omega\sqrt{\varepsilon\mu}=\frac{\omega}{c}\sqrt
{\varepsilon_{r}\mu_{r}}$ is the wave number. Applying the operator $D$ to the
functions $\overrightarrow{\varphi}$ and $\overrightarrow{\psi}$ one can see
that $\overrightarrow{\varphi}$ satisfies the equation
\begin{equation}
(D-\kappa)\overrightarrow{\varphi}=0, \label{D-phi}%
\end{equation}
and $\overrightarrow{\psi}$ satisfies the equation
\begin{equation}
(D+\kappa)\overrightarrow{\psi}=0. \label{D+psi}%
\end{equation}
Solutions of (\ref{D-phi}) and (\ref{D+psi}) are called the Beltrami fields
(see, e.g., \cite{Lakhtakia}).

\section{The relation}

In the preceding sections it was shown that the Dirac equation (\ref{Diract})
is equivalent to the equation $D_{\overrightarrow{\alpha}}f=0$ with
$\overrightarrow{\alpha}=-\frac{1}{\hbar}(i\frac{\mathcal{E}}{c}i_{1}%
+mci_{2})$ and the Maxwell equations (\ref{Mh1})-(\ref{Mh4}) are equivalent to
the pair of quaternionic equations $D_{-\kappa}\overrightarrow{\varphi}=0$ and
$D_{\kappa}\overrightarrow{\psi}=0$. Now we will show a simple relation
between these objects. Suppose that%
\begin{equation}
\kappa^{2}=\overrightarrow{\alpha}^{2}. \label{rel}%
\end{equation}
Let us introduce the following operators of multiplication%

\[
P^{\pm}:=\frac{1}{2\kappa}M^{\kappa\pm\overrightarrow{\alpha}}.
\]
It is easy to verify that they are mutually complementary and orthogonal
projection operators, and the following equality is valid \cite{KSbook}
\begin{equation}
D_{\overrightarrow{\alpha}}=P^{+}D_{\kappa}+P^{-}D_{-\kappa}. \label{relop}%
\end{equation}
Moreover, as $P^{\pm}$ commute with $D_{\pm\kappa}$, we obtain that any
solution of (\ref{Dirac}) is uniquely represented as follows%
\[
f=P^{+}\psi+P^{-}\varphi,
\]
where $\varphi$ and $\psi$ are solutions of (\ref{D-phi}) and (\ref{D+psi})
respectively but in general can be full quaternions not necessarily purely
vectorial. In particular, we have that
\[
f=P^{+}(i\omega\varepsilon\overrightarrow{E}+\kappa\overrightarrow{H}%
)+P^{-}(-i\omega\varepsilon\overrightarrow{E}+\kappa\overrightarrow{H})=
\]%
\[
i\omega\varepsilon(P^{+}-P^{-})\overrightarrow{E}+\kappa(P^{+}+P^{-}%
)\overrightarrow{H}=
\]%
\[
\frac{i\omega\varepsilon}{\kappa}\overrightarrow{E}\cdot\overrightarrow
{\alpha}+\kappa\overrightarrow{H}
\]
is a solution of (\ref{Dirac}) if $\overrightarrow{E}$ and $\overrightarrow
{H}$ are solutions of (\ref{Mh1})-(\ref{Mh4}).

It should be noticed that (\ref{relop}) works in both directions. We have%

\[
D_{\kappa}=P^{+}D_{\overrightarrow{\alpha}}+P^{-}D_{-\overrightarrow{\alpha}}
\]
and%
\[
D_{-\kappa}=P^{-}D_{\overrightarrow{\alpha}}+P^{+}D_{-\overrightarrow{\alpha}%
}.
\]
The fact that the Maxwell system reduces to equations (\ref{D-phi}) and
(\ref{D+psi}), where the functions $\overrightarrow{\varphi}$ and
$\overrightarrow{\psi}$ are purely vectorial provokes the natural question
whether it had any sense to consider full quaternions $\varphi$ and $\psi$ and
hence four-component vectors $E$ and $H$ or the nature definitely eliminated
their scalar parts. Some arguments supporting the idea of nonzero scalar parts
can be found, for example, in \cite{Dvoeglazov}.

As we have seen equality (\ref{relop}) is valid under the condition
(\ref{rel}). Let us analyze this condition. Note that
\[
\overrightarrow{\alpha}^{2}=-<\overrightarrow{\alpha},\overrightarrow{\alpha
}>=\frac{1}{\hbar^{2}}(\frac{\mathcal{E}^{2}}{c^{2}}-m^{2}c^{2}).
\]
Thus (\ref{rel}) has the form%
\begin{equation}
\kappa^{2}=\frac{1}{\hbar^{2}}(\frac{\mathcal{E}^{2}}{c^{2}}-m^{2}c^{2})
\label{surpr}%
\end{equation}
or equivalently%

\[
(\hbar\omega)^{2}\varepsilon_{r}\mu_{r}=\mathcal{E}^{2}-m^{2}c^{4}.
\]
From this equation in the case $\varepsilon_{r}=\mu_{r}=1$, that is for a
vacuum, using the well known in quantum mechanics relation between the
frequency and the impulse: $\hbar\omega=pc$ we obtain the equality
\begin{equation}
\mathcal{E}^{2}=p^{2}c^{2}+m^{2}c^{4}. \label{fundeq}%
\end{equation}
In general, if in (\ref{surpr}) we formally use the de Broglie equality
$p=\hbar\kappa$, we again obtain the fundamental relation (\ref{fundeq}).

Thus relation (\ref{relop}) between the Dirac operator and the Maxwell
operators is valid if the condition (\ref{surpr}) is fulfilled which quite
surprisingly is in agreement with (\ref{fundeq}).

\end{document}